# Fabrication of astronomical X-ray reflection gratings using thermally activated selective topography equilibration (TASTE)

Jake A. McCoy [a]

The Pennsylvania State University, Department of Astronomy and Astrophysics, 525 Davey Laboratory, University Park, Pennsylvania 16802

Randall L. McEntaffer

The Pennsylvania State University, Department of Astronomy and Astrophysics, 505 Davey Laboratory, University Park, Pennsylvania 16802

Chad M. Eichfeld

The Pennsylvania State University, Materials Research Institute, N-153 Millenium Science Complex, University Park, Pennsylvania 16802

[a] Electronic mail: jam1117@psu.edu

Thermally activated selective topography equilibration (TASTE) enables the creation of 3D structures in resist using grayscale electron-beam lithography followed by a thermal treatment to induce a selective polymer reflow. A blazed grating topography can be created by reflowing repeating staircase patterns in resist into wedge-like structures. Motivated by astronomical applications, such patterns with periodicities 840 nm and 400 nm have been fabricated in 130 nm-thick PMMA using TASTE to provide a base for X-ray reflection gratings. A path forward to integrate this alternative blazing technique into grating fabrication recipes is discussed.

## I. INTRODUCTION

Common practice of electron-beam lithography (EBL) is to create bi-level structures in a film of polymeric resist, where high-energy electrons are used to alter local average polymer molecular weight $M_w$. For positive-tone resists such as poly(methyl methacrylate) (PMMA), these binary patterns are produced when $M_w$ is reduced upon electron exposure[1], causing the polymer to be etched away during wet development. In contrast, multi-level structures can be patterned in resist using grayscale EBL[2] (GEBL). Here, dose-modulated electron exposure imparts a range of locally modified $M_w$ in the resist, resulting in correspondingly varying developer solubility. Using a timed development, multi-level structures can be attained as resist is etched down to a thickness that depends on local $M_w$. However, 3D topographies can only be approximated using multi-level steps in GEBL; thermally activated selective topography equilibration[3,4,5] (TASTE) is a technique based on the principle that reduced $M_w$ from electron exposure also results in a lowered polymer glass-liquid transition temperature $T_g$, above which resist is able to flow according to a time-dependent visco-elastic creep process[6]. Using TASTE, a wafer featuring a GEBL pattern can be heated such that electron-exposed resist is allowed to equilibrate in a molten state while unexposed resist stays in its glass state. In this way, micron-scale 3D patterns such as sloped structures have been fabricated in PMMA[5], ZEP520[7] and mr-PosEBR[8].

Next-generation reflection gratings for astronomical X-ray spectroscopy demand custom architectures to enable sensitive observations with high spectral resolving power ($R = \lambda/\Delta\lambda$) in the soft X-ray regime ($\lambda = 6 - 62$ Å). Spectrometers for this application typically consist of planar gratings that are stacked and aligned into modular arrays to intercept the radiation coming to a focus in a Wolter-I telescope, where a detector, such as a CCD camera, placed at the focal plane is used to image the dispersed spectrum[9,10,11]. To achieve high $R$, gratings with periodicities on the order of hundreds of nanometers require precise variable-line-space groove layouts to match the focal length of the telescope[12] and blazed topographies to maximize diffraction efficiency for a particular band of interest. Standard approach is to define a groove layout in resist using lithography followed by processing to provide a blaze, such as wet anisotropic etching or directional ion milling. As an alternative to these methods, TASTE can be used to pattern a blazed grating structure directly in resist over a custom layout. This paper explores the feasibility of using TASTE to generate blazed topographies suitable for X-ray grating spectroscopy and discusses several options to produce functional reflection gratings from such a template.

## II. EXPERIMENTAL

Fabrication of blazed grating patterns with periodicities 840 nm and 400 nm in ~130 nm-thick PMMA is described in the following: carried out at the Nanofabrication Laboratory at the Pennsylvania State University (PSU) Materials Research Institute, methods consist of GEBL to create multi-level staircase structures, spectroscopic ellipsometry (SE) for resist contrast data collection, and thermal reflow by hotplate to smooth exposed resist into a sawtooth topography. Patterns were examined using atomic force microscopy (AFM; Bruker Icon) at the PSU Materials Characterization Laboratory.

## A. Grayscale lithography

Practicing GEBL requires knowledge of resist contrast to map electron dose to remaining film thickness for a given development recipe. Therefore, it is important to control resist conditions throughout process development. All patterns were created in PMMA (MicroChem Corp.) with $M_w = 950$ kg mol$^{-1}$. For each test sample, a clean, polished 4" Si wafer, purchased from Virginia Semiconductor, Inc., was dehydration-baked at 180°C prior to spin coating. The resist, diluted 3% in Anisole, was spun at 3 krpm using a dynamic dispense to yield a 130 nm thickness as measured by SE. Following coating, wafers were baked again at 180°C for 3 minutes to remove residual Anisole from the resist. Electron exposure was performed at 100 kV using a Raith EBPG5200 system with data preparation facilitated using the Layout BEAMER software package (GenISys GmbH). Samples were developed at room temperature for 2 minutes in a 1:1 mixture of methyl isobutyl ketone (MIBK) and isopropyl alcohol (IPA), followed by a 30-second IPA bath and a nitrogen blow-dry. To allow solvent-induced gel and resist swelling to subside, test samples were not characterized using SE or AFM until ~24 hours post development.

Resist contrast data collection requires the fabrication and analysis of a pattern designed to probe resist thickness as a function of electron dose. The layout of this pattern, shown in Fig. 1, consists of alignment markers for automated SE data acquisition (large crosses) and a 5 by 5 array of 250 µm-wide squares dosed to a range of values using BEAMER. Starting with 50 µC/cm$^2$, the pattern was exposed using a 30 nA beam with a 200 µm aperture to dose each square in 7% increments. To measure the resist thickness in each square, SE data were gathered at the center of each square using a focused M-2000 ellipsometer (J.A. Woollam). Using Woollam's CompleteEASE software, data were fit to a transparent film model for PMMA over 450 nm<$\lambda$<1000 nm using Cauchy's equation:

$$n(\lambda) = A + B/\lambda^2 + C/\lambda^4 \quad (1)$$

where $n$ is index of refraction and $A$, $B$ and $C$ are fit parameters. As As $n$ is expected to change with electron exposure, $A$ and $B$ were fit for each measurement, keeping $C$=0. From these fits, resist thickness was extracted and plotted against electron dose, shown in Fig. 2. This resist contrast curve served as the basis for GEBL in 130 nm-thick PMMA under these conditions.

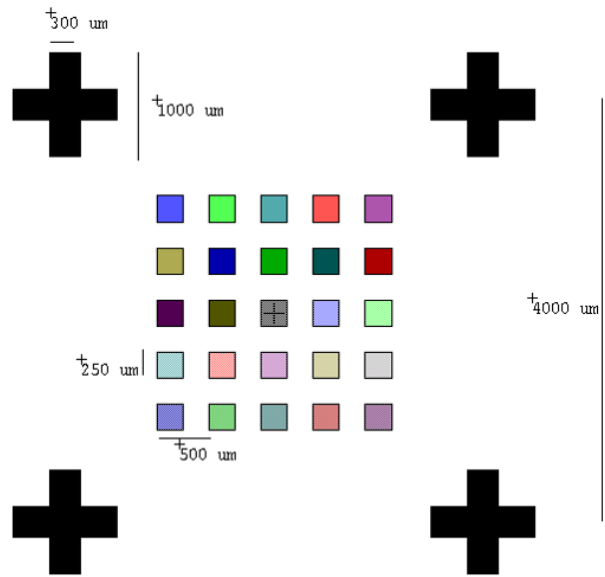

FIG. 1. (Color online) A layout is defined for resist contrast data collection using Tanner L-Edit software where an array of 25 squares, each 250 µm in width, is assigned to a range of values using Layout BEAMER.

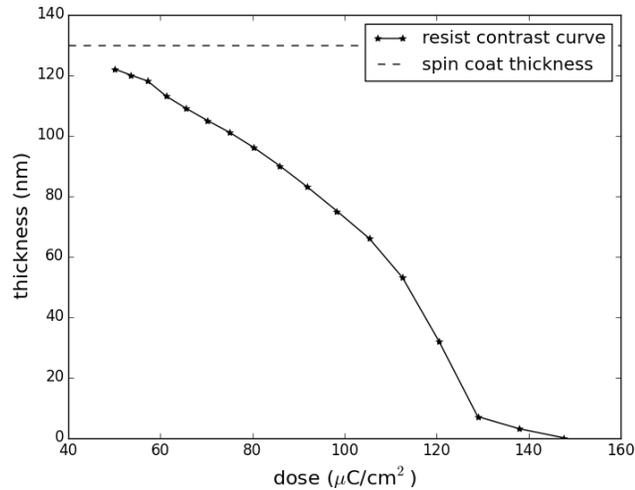

FIG. 2. Resist contrast for 130 nm-thick PMMA developed at room temperature using 1:1 MIBK/IPA for 2 minutes and IPA for 30 s: the center of each square of the pattern shown in Fig. 1 is measured using spectroscopic ellipsometry and a film thickness is extracted. Resist contrast is plotted as remaining resist thickness as a function of electron dose.

Standard approach for GEBL process development was to draw computer aided design (CAD) for multi-level layouts, mapping design layer to intended resist thickness. Though a contrast curve gives remaining resist thickness as a function of electron dose by definition, proximity effect correction[13] (PEC) must be carried out to attain accurate GEBL structures. This was handled using BEAMER's 3DPEC algorithm[14], which uses input resist contrast data and a point spread function for electron backscattering in silicon, determined through Monte Carlo simulation, to calculate how dose is distributed for PEC. The result is a relative dose-corrected layout fractured into 100 layers; this is exported to EBPG5200 machine format as 100 pattern generator (PG) shapes. Patterns exposed in this way used a 1 nA beam current, a 200 µm aperture and a 10 nm beam step size (BSS).

Due to the added overhead associated with switching between PG shapes, an alternative approach utilizing EBPG sequencing was also pursued. Unique to EBPG systems, sequencing is a set of commands that defines a series of lines and beam jumps to be executed as a custom, subfield-sized PG shape. This shape can be patterned over an arbritrary area using an array of subfield-sized rectangles defined in CAD. Though sequencing is limited to lines and jumps at a single user-input electron dose, dose modulation can be emulated using overlapping beams (see Fig. 3). In this way, GEBL patterns were produced by generating sequencing code for a 4 µm-large custom PG shape to approximate the dose-corrected layout from 3DPEC. However, at the edges of a grating area, extra dose is needed for PEC; this is not compensating for using this approach. Patterns exposed in this way used beam currents ~25 nA, a 400 µm aperture and a 40 nm BSS. The input electron dose was determined through analysis of dose test arrays and the highest possible beam current (considering limitations imposed by the 25 MHz EBPG5200 clock frequency) was used to write each pattern.

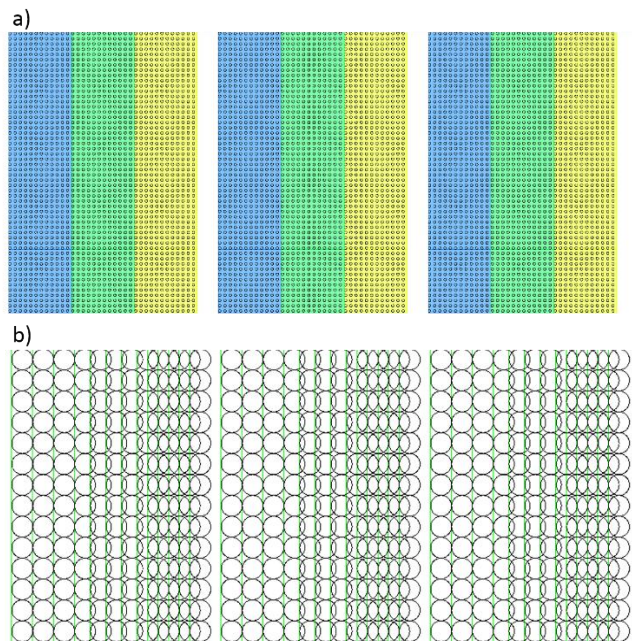

FIG. 3. (Color online) Two approaches to pattern generation for grayscale electron-beam lithography: a) Experimentally determined resist contrast data are input into Layout BEAMER where the 3DPEC algorithm is used to generate a dose-corrected layout. Each dose is a different pattern generator shape, represented here with colorscale. b) Code for EBPG5200 sequencing defines a custom subfield-sized pattern generator shape consisting of a series of lines and beam jumps. Lines are programmed to overlap to varying degrees to emulate dose modulation achieved using the 3DPEC approach.

Using these two GEBL approaches, multi-level staircase structures with 840 nm and 400 nm periodicities were patterned in resist. In general, the width of staircase steps were designed to be comparable to the spin coat thickness of 130 nm. A series of GEBL test samples were written over 0.5 mm by 2 mm areas featuring the following patterns:

A) 6-level staircase: each step is 140 nm wide to give a 840 nm periodicity

B) 4-level staircase: each step is 100 nm wide to give a 400 nm periodicity

C) 4-level staircase: width of the top step is reduced to 40 nm; remaining levels are 120 nm wide to give a periodicity of 400 nm

D) Emulation of pattern C attained through EBPG5200 sequencing

In principle, the staircase steps intermediate between the top step and the exposed substrate will equilibrate into sloped surfaces using an optimized thermal treatment. The width of the top step in patterns C and D was reduced in an effort to minimize flat area atop groove structures in the final product. Ideally, grooves are sharp, triangular sawtooth facets to provide an effective blazed grating response.

## B. Thermal reflow

All thermal reflow experimentation was carried using an automated hotplate on a resist stabilization system built by Fusion Semiconductor. The tool, which accepts 4" wafers, was used to heat test samples in a controllable and reproducible way. A series of identical GEBL samples were fabricated and reflowed under different conditions to probe the temperature-time parameter space for TASTE. Using a short heating time (20 s), samples were treated using s range of temperatures around the expected $T_g$ for PMMA (110-130°C). Conversely, samples were heated for different durations (20 s, 60 s, 120 s) while holding temperature constant at 120°C. Each sample was allowed to cool on the cassette rack of the automated hotplate tool after heating.

# III. RESULTS

Untreated GEBL structures in PMMA for patterns A-D are seen under AFM in Fig. 4. Using Brukers' PeakForce Tapping™ mode, patterns were scanned over 2 µm at 512 samples per line to give a 3.9 nm AFM pixel size. In patterns A and B, staircase steps are equal in width and the overall structure height is comparable to the initial spin coat thickness as measured by AFM. However, due to the narrowed width of the top step in patterns C and D, the overall structure height is reduced to about 75% of the original film thickness. Patterns A-C were fabricated using the BEAMER 3DPEC approach, where a 1 mm$^2$ area is exposed in ~20 minutes using the EBPG5200. In contrast, pattern D over the same area can be written using EBPG5200 sequencing in ~1 minute. Although pattern C exhibits higher noise than pattern D under AFM, the end result after thermal reflow should be comparable. For this reason, the EBPG5200 sequencing approach is valuable for patterning GEBL over large areas, where it is imperative for write time to be minimized.

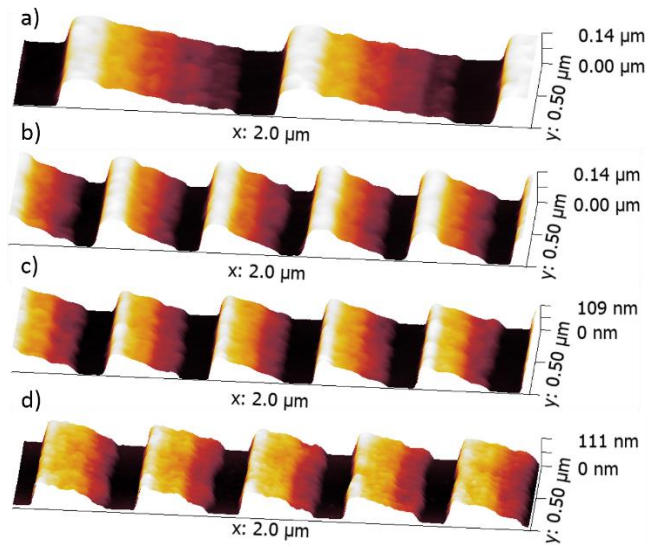

FIG. 4. (Color online) Grayscale electron-beam lithography (GEBL) in 130 nm-thick PMMA examined under atomic force microscopy. Patterns A-C were attained through the standard 3DPEC approach while pattern D is a result from the EBPG5200 sequencing approach: a) Pattern A: GEBL with 6 levels where each are 140 nm wide to give a 840 nm periodicity. b) Pattern B: GEBL with 4 levels where each are 100 nm wide to give a 400 nm periodicity. c) Pattern C: GEBL with 4 levels where the width of the top level is reduced to 40 nm; remaining levels are 120 nm wide to give a periodicity of 400 nm. d) Pattern D: same as pattern C but attained using EBPG sequencing.

Thermal reflow results for pattern A are shown in Fig. 5. This includes six AFM scans: three samples treated at 110°C, 120°C and 130°C keeping heating time constant at 20 s, and conversely, three samples heated with durations of 20 s, 60 s and 120 s holding temperature at 120°C. Analogous AFM scans of pattern C are shown in Fig. 6. These results demonstrate the principle of TASTE: stepped surfaces of resist begin to smooth into inclines with an appropriate thermal treatment. However, only a small sample of the temperature-time parameter space for TASTE has been explored. Reflow temperature should be optimized such that intermediate steps are able to flow while the top step is unaffected; this is to achieve a sharp sawtooth by avoiding groove rounding effects. Meanwhile, heating duration should also be optimized to allow flowing resist to equilibrate into smooth, sloped surfaces. Preffered temperature-time parameters will depend on the intitial GEBL structure. Shown in Fig. 5, a small reflow effect at 110°C is seen in the intermediate steps of pattern A. The effect becomes more pronounced at 120°C while at 130°C, the entire structure, including the top step, is affected. At 120°C, the intermediate steps become increasingly smooth as reflow time reaches 120 s, at which point equilibration is not yet reached. This suggests that the optimum temperature for TASTE of pattern A is close to 120°C while the optimum reflow duration may be greater than 120 s. Shown in Fig. 6, a similar effect is observed for pattern C. However, the optimum reflow temperature here may be slightly lower as the narrowed top step is expected to have a reduced $M_w$ relative to that of pattern A. This is evidenced by the ~25% reduction in structure height relative to the original film thickness and the slight rounding effects observed at 120°C for a 120 s duration, which reflect a slightly lowered $T_g$ as compared to pattern A.

Although not exhaustive, these results show that a process window exists to transform a staircase structure into the desired wedge shape using thermal reflow. Future investigations for TASTE will include repeating these hotplate experiments over matrices of finer temperature increments and heating longer durations. For example, for pattern A, this might consist of treating samples at 116°C, 118°C, 120°C, 122°C and 124°C for durations of 100 s, 110 s, 120 s, 130 s and 140 s. As the optimum reflow temperature for pattern C is expected to be slightly reduced, an appropriate test matrix for this pattern might be 112°C, 114°C, 116°C, 118°C and 120°C over similar durations. If the desired results are not found within these parameter spaces, lower temperatures with even longer heating durations may be explored.

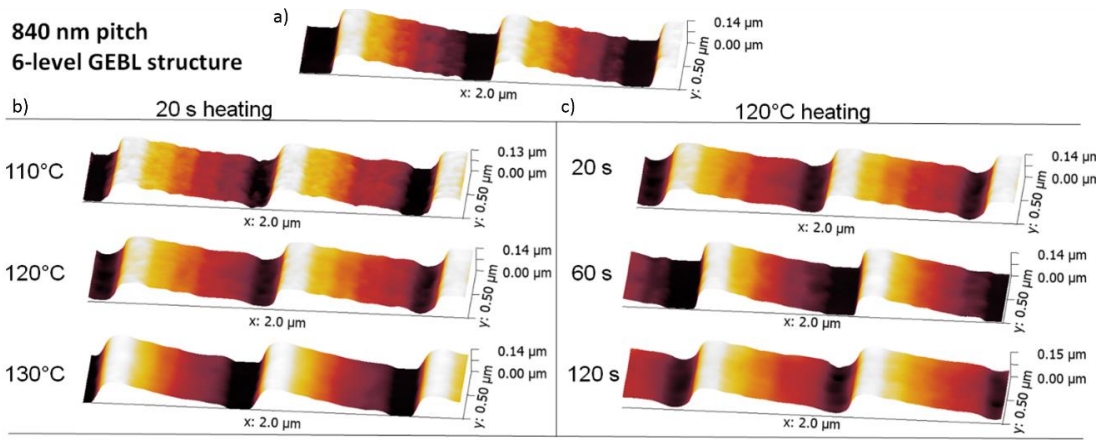

FIG. 5. (Color online) Thermal reflow results for grayscale electron-beam lithography in 130 nm-thick PMMA with 6 levels: each are 140 nm wide to give a 840 nm periodicity. a) Intitial GEBL pattern. b) Thermal reflow at various temperatures with constant heating duration. c) Thermal reflow at constant temperature with various heating durations.

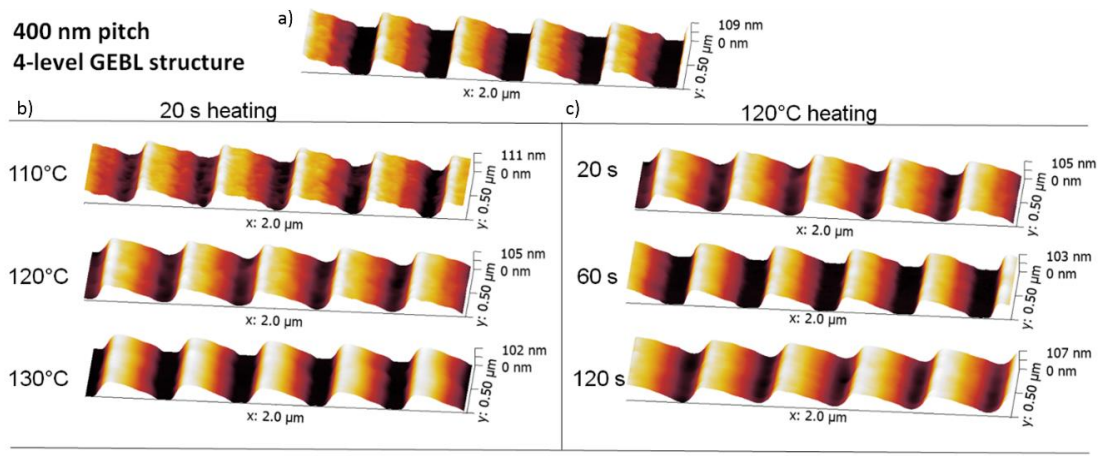

FIG. 6. (Color online) Thermal reflow results for grayscale electron-beam lithography in 130 nm-thick PMMA with 4 levels: width of the top level is reduced to 40 nm; remaining levels are 120 nm wide to give a periodicity of 400 nm. a) Intitial GEBL pattern. b) Thermal reflow at various temperatures with constant heating duration. c) Thermal reflow at constant temperature with various heating durations.

In addition to the overall topography achieved using TASTE, surface roughness and EBL field stitching are of importance in grating fabrication. The TASTE results have sloped surfaces with RMS roughness of a few nanometers as measured by AFM; ideally, this roughness would be reduced with better optimized thermal reflow conditions. For patterns written using a 300 µm mainfield for the EBPG5200, field stitch boundaries are observable post reflow under AFM (see Fig. 7). Along the groove direction, the field stitch manifests itself as added facet roughness, which may be improved under the appropriate reflow conditions. Controlling field stitching along the cross-groove direction is also important to avoid irregularities in groove spacing. These parameters will be explored in more detail in a forthcoming publication.

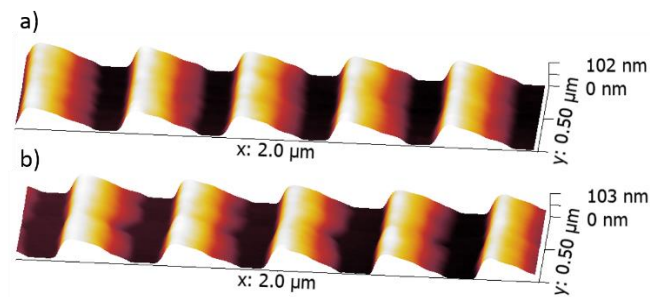

FIG. 7. (Color online) Grayscale electron-beam lithography in 130 nm-thick PMMA with 4 levels thermally treated at 120°C for 60 s: a) near the center of a 300 µm EBPG mainfield and b) at a field stitch boundary.

## IV. DISCUSSION

The blazed topography of choice depends on the grating geometry employed; a general case is shown in Fig. 8 where groove spacing $d$ determines the locations of diffracted orders for a wavelength $\lambda$ through the generalized grating equation:

$$\sin(\alpha) + \sin(\beta) = n\,\lambda/d\,\sin(\gamma) \quad (2)$$

Here, $\alpha$ is the polar incidence angle, $\beta$ is the polar diffracted angle for a diffracted order $n$ and $\gamma$ is the half-cone opening angle between the incident beam and the groove direction. Moreover, for a grating with blaze angle $\delta$, radiation is preferentially diffracted to $\beta = 2\,\delta - \alpha$ and the blaze wavelength for order $n$ is given by:

$$\lambda_{blaze} = d\,\sin(\gamma)(\sin(\alpha) + \sin(2\,\delta - \alpha))/n. \quad (3)$$

X-ray reflection gratings are typically designed for use at grazing incidence angles where the angle of incident radiation relative to the plane defined by the blazed groove facet is on the order of a few degrees. This angle $\zeta$ is given by:

$$\sin(\zeta) = \sin(\gamma)\cos(\delta - \alpha). \quad (4)$$

Gratings are commonly used in an in-plane mount where the groove direction is perpendicular to incoming radiation and hence $\sin(\gamma) = 1$. This geometry necessitates shallow $\delta$, large $\alpha$ and $d$ on the order of several hundred nanometers. Another option is to orient grating grooves to be quasi-parallel to incoming radiation so that $\gamma \lessapprox 2°$. This extreme off-plane mount has several attractive features including minimal groove shadowing and the ability to tightly pack gratings into an array[15]. In this geometry, $d$ on the order of a few hundred nanometers or less is required while there is more freedom for $\delta$ and $\alpha$ (typically, Littrow configuration is used: $\alpha = \beta = \delta$). Mid to steep $\delta$ is desirable to preferentially disperse radiation to high $n$, where large dispersion can be attained.

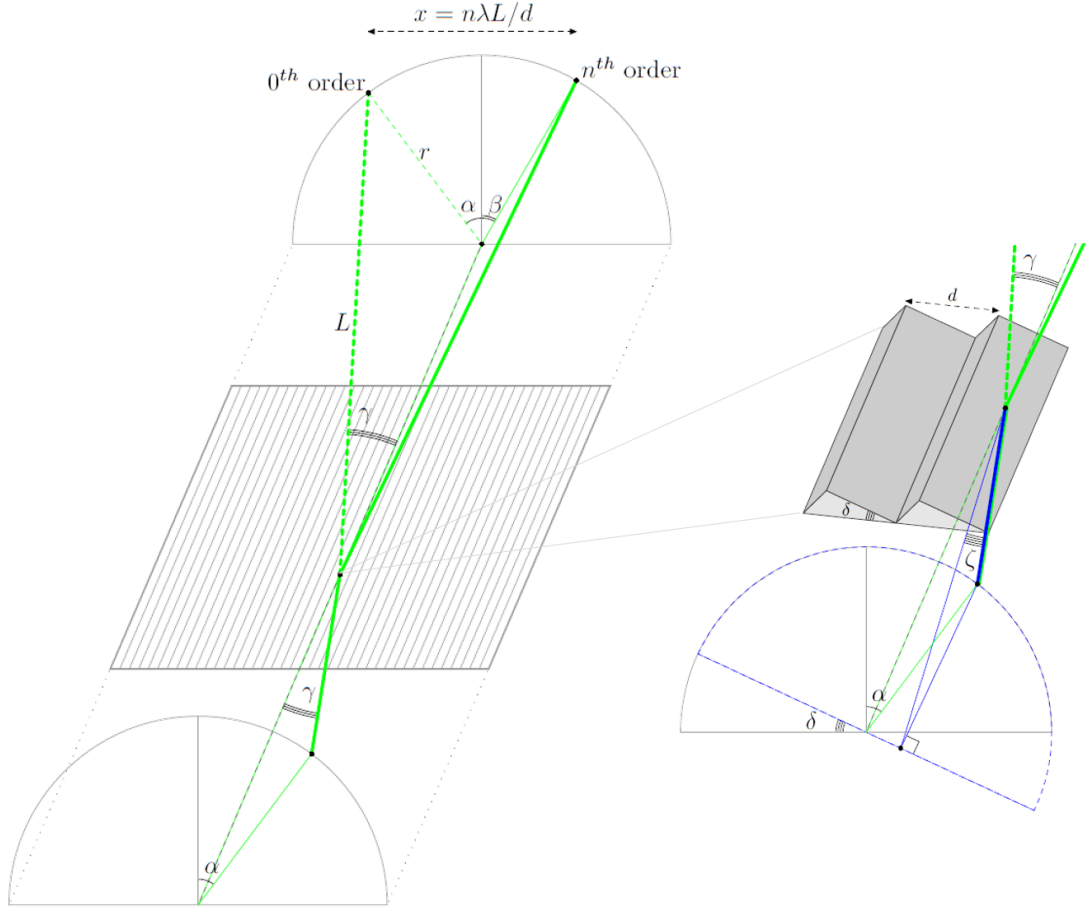

FIG. 8. (Color online) Grating geometry: many applications utilize an in-plane geometry where $\sin(\gamma) = 1$ but off-plane geometries with $\gamma \lessapprox 2°$ are beneficial for grazing-incidence spectrometers with modular grating arrays. For a given throw $L$, diffracted orders lay on an arc with radius $r = L \sin(\gamma)$ and grating dispersion $x$ is proportional to wavelength $\lambda$, order number $n$ and inversely proportional to the groove spacing $d$. Gratings are blazed to a sawtooth profile with facet angle $\delta$. The angle at which radiation is incident on blazed groove facets, $\zeta$, must be kept small for grazing-incidence applications, regardless of the geometry employed.

Results show a path forward for fabricating X-ray reflection gratings using TASTE. Patterns at $d = 840$ nm yield $\delta \sim 10°$ and are more suitable for in-plane applications while patterns at $d = 400$ nm provide a base for off-plane gratings with $\delta \sim 25°$. However, such a topography in resist on its own is not ready to be used directly as an X-ray reflection grating. Instead, the pattern should be stabilized and coated with a metal for soft X-ray reflectivity; owing to small X-ray attenuation depth at grazing incidence, only 10-20 nm of deposition is required. As untreated PMMA is susceptible to damage from physical vapor deposition processes, extra processing is needed to functionalize the pattern for spectroscopy. This may include one of the following options:

1) Transfer the pattern into Si with a dry etch and coat the substrate with a metal to produce a single grating

2) Cross-link PMMA through UV exposure and coat the cured resist with a metal to produce a single grating

3) Transfer the pattern into Si with a dry etch, use the etched substrate as a direct stamp for nanoimprint lithography[16,17] (NIL), and coat the resulting imprints with a metal to produce many grating replicas

4) Transfer the pattern into Si with a dry etch, use the etched substrate for the construction of a composite stamp for substrate conformal imprint lithography[18] (SCIL), and coat the resulting imprints with a metal to produce many grating replicas

5) Construct a composite stamp for SCIL directly from the resist and coat the resulting imprints with a metal to produce many grating replicas

The path forward for grating fabrication will depend on the application. The TASTE results presented here have sloped surfaces with RMS roughness $\sigma$ of a few nanometers as measured by AFM. Because the Rayleigh scattering criterion for a smooth surface demands $4\pi\sigma \sin(\zeta) < 0.1\ \lambda$, dry etch induced roughness is a concern. However, transferring the pattern into Si opens up the door for resist modification, such as sequential infiltration synthesis to increase the etch selectivity of PMMA[19]; this would allow the creation of groove facets with a steepened $\delta$ relative to what is directly patterned with TASTE. On the other hand, if many gratings are required to populate modular grating arrays, rather than directly fabricating each using TASTE it is more cost-effective to fabricate a master grating and produce replicas using imprint lithography. For large-area gratings, it is beneficial to utilize the EBPG5200 sequencing approach, where EBL write times are reduced by a factor of ~20 relative to the BEAMER 3DPEC approach. Moreover, large-area patterns pose challenges for imprint lithography: for NIL, the quality of the contact between the mold and the substrate requires careful monitoring to ensure an even contact distribution across the entire grating surface to avoid defects borne out of trapped air pockets during imprinting. Some of these problems may be evaded using SCIL, where a flexible composite stamp is able to conform globally to the bow of the substrate and locally around particulate contaminants that may be present during imprinting.

## V. SUMMARY AND CONCLUSIONS

Using GEBL, staircase strucutes with periodicity $d = 840$ nm and $d = 400$ nm were fabricated in 130 nm-thick PMMA and through a thermal treatment, these patterns were smoothed into 3D wedge-like structures to generate blazed grating topographies. Further optimization of the TASTE process for large-area, replicated gratings will be presented in forthcoming publications. These gratings will also be tested for spectral resolving power and diffraction efficiency to determine experimentally if TASTE can be used to push the state-of-the-art for X-ray diffraction gratings.

## ACKNOWLEDGMENTS

This research used resources of the Nanofabrication Laboratory and the Materials Characterization Laboratory at the Penn State Materials Research Institute and was supported by a NASA Space Technology Research Fellowship (NSTRF) under grant NNX16AP92H.

FIG. 1. (Color online) A layout is defined for resist contrast data collection using Tanner L-Edit software where an array of 25 squares, each 250 µm in width, is assigned to a range of values using Layout BEAMER.

FIG. 2. Resist contrast for 130 nm-thick PMMA developed at room temperature using 1:1 MIBK/IPA for 2 minutes and IPA for 30 s: the center of each square of the pattern shown in Fig. 1 is measured using spectroscopic ellipsometry and a film thickness is extracted. Resist contrast is plotted as remaining resist thickness as a function of electron dose.

FIG. 3. (Color online) Two approaches to pattern generation for grayscale electron-beam lithography: a) Experimentally determined resist contrast data are input into Layout BEAMER where the 3DPEC algorithm is used to generate a dose-corrected layout. Each dose is a different pattern generator shape, represented here with colorscale. b) Code for EBPG5200 sequencing defines a custom subfield-sized pattern generator shape consisting of a series of lines and beam jumps. Lines are programmed to overlap to varying degrees to emulate dose modulation achieved using the 3DPEC approach.

FIG. 4. (Color online) Grayscale electron-beam lithography (GEBL) in 130 nm-thick PMMA examined under atomic force microscopy. Patterns A-C were attained through the standard 3DPEC approach while pattern D is a result from the EBPG5200 sequencing approach: a) Pattern A: GEBL with 6 levels where each are 140 nm wide to give a 840 nm periodicity. b) Pattern B: GEBL with 4 levels where each are 100 nm wide to give a 400 nm periodicity. c) Pattern C: GEBL with 4 levels where the width of the top level is reduced to 40 nm; remaining levels are 120 nm wide to give a periodicity of 400 nm. d) Pattern D: same as pattern C but attained using EBPG sequencing.

FIG. 5. (Color online) Thermal reflow results for grayscale electron-beam lithography in 130 nm-thick PMMA with 6 levels: each are 140 nm wide to give a 840 nm periodicity. a) Intitial GEBL pattern. b) Thermal reflow at various temperatures with constant heating duration. c) Thermal reflow at constant temperature with various heating durations.

FIG. 6. (Color online) Thermal reflow results for grayscale electron-beam lithography in 130 nm-thick PMMA with 4 levels: width of the top level is reduced to 40 nm; remaining levels are 120 nm wide to give a periodicity of 400 nm. a) Intitial GEBL pattern. b)

Thermal reflow at various temperatures with constant heating duration. c) Thermal reflow at constant temperature with various heating durations.

FIG. 7. (Color online) Grayscale electron-beam lithography in 130 nm-thick PMMA with 4 levels thermally treated at 120°C for 60 s: a) near the center of a 300 µm EBPG mainfield and b) at a field stitch boundary.

FIG. 8. (Color online) Grating geometry: many applications utilize an in-plane geometry where $\sin(\gamma) = 1$ but off-plane geometries with $\gamma \lessgtr 2°$ are beneficial for grazing-incidence spectrometers with modular grating arrays. For a given throw $L$, diffracted orders lay on an arc with radius $r = L \sin(\gamma)$ and grating dispersion $x$ is proportional to wavelength $\lambda$, order number $n$ and inversely proportional to the groove spacing $d$. Gratings are blazed to a sawtooth profile with facet angle $\delta$. The angle at which radiation is incident on blazed groove facets, $\zeta$, must be kept small for grazing-incidence applications, regardless of the geometry employed.